\documentclass[sigconf, nonacm]{acmart}

\usepackage{soul}
\usepackage{enumitem}
\usepackage{makecell}
\usepackage{hhline}
\usepackage{tabularx}
\usepackage{algorithm}
\usepackage{algorithmic}
\usepackage{graphicx}
\usepackage{subcaption}
\usepackage{listings}
\usepackage{amsmath}
\usepackage{balance}

\usepackage[colorinlistoftodos]{todonotes}
\presetkeys{todonotes}{inline}{}

\newcolumntype{V}{>{\centering\arraybackslash}X}

\begin{document}

\title{TreeCat: Standalone Catalog Engine for Large Data Systems}

\author{Keonwoo Oh}
\affiliation{%
    \institution{University of Maryland}
    \city{College Park}
    \state{Maryland}
    \country{USA}}
\email{koh3@umd.edu}

\author{Pooja Nilangekar}
\affiliation{%
    \institution{University of Maryland}
    \city{College Park}
    \state{Maryland}
    \country{USA}
}
\email{poojan@umd.edu}

\author{Amol Deshpande}
\affiliation{%
    \institution{University of Maryland}
    \city{College Park}
    \state{Maryland}
    \country{USA}
}
\email{amol@umd.edu}

\begin{abstract}

With ever-increasing volume and heterogeneity of data, advent of new specialized compute engines, and demand for complex use cases, large-scale data systems require a performant catalog system that can satisfy diverse needs. We argue that existing solutions, including recent lakehouse storage formats, have fundamental limitations and that there is a strong motivation for a specialized database engine, dedicated to serve as the catalog. We present the design and implementation of {\em TreeCat}, a database engine that features a hierarchical data model with a path-based query language, a storage format optimized for efficient range queries and versioning, and a correlated scan operation that enables fast query execution. A key performance challenge is supporting concurrent read and write operations from many different clients while providing strict consistency guarantees. To this end, we present a novel MVOCC (multi-versioned optimistic concurrency control) protocol that guarantees serializable isolation. We conduct a comprehensive experimental evaluation comparing our concurrency control scheme with prior techniques, and evaluating our overall system against Hive Metastore, Delta Lake, and Iceberg.

\end{abstract}

\maketitle

\section{Introduction}
\label{sec:intro}

The catalog is an important component of a database management system that is responsible for storing and maintaining metadata such as the properties of logical entities like databases, tables, views, triggers, and
schemas; mapping of these entities to the underlying physical storage; and access control information. Metadata operations are in the hot path of most, if not all, database operations. Take, for example, the
query execution process of a relational database. The table schema is necessary to first semantically validate the query. Once the query is validated, the query planner uses various metadata, including table schema, storage
information, and statistics, to optimize and build a physical query plan that is executed by the execution engine (which usually needs access to the metadata as well). Without a high-performance catalog that provides fast access to metadata, a database management system cannot operate efficiently.

In a monolithic data system, the catalog is typically tightly coupled with the rest of the system, which works for one or a few data systems. But in a typical enterprise setting, organizations often use many specialized engines, each serving a different function, since no single one-size-fits-all system can meet the requirements of all use cases. Meanwhile, there has been a growing trend toward disaggregation of large-scale data systems into smaller subsystems. This includes the separation of compute and storage layers, the increasing adoption of common table formats (parquet, ORC), and the use of ``connectors" to link different systems. These efforts aim to achieve a composable data stack, enabling organizations to store data in a shared storage layer and choose from multiple compute engines. However, as long as each system maintains its own catalog such that the metadata about the ``managed" data remain separate, different systems cannot easily share data even if it is stored in the same storage system. {\em For data management systems to be truly composable, the catalog functionality has to be disaggregated from the rest of the system.} This issue is now widely recognized in the data systems industry, and there are multiple active projects, including lakehouse storage formats~\cite{deltalake,apacheiceberg,apachehudi}, and, most recently, catalog services~\cite{iceberg_rest,unity,openhouse,polaris}, mainly driven by the industry.

We further argue that not only should the catalog functionality exist as a standalone service, but also that it should be unified and not span multiple (semi-)autonomous systems or storage modules. 
As we show, unified architecture not only leads to significantly reduced latencies, but also reduces duplication of metadata across different data engines, and enables multi-table transactions as well as other complex operations on metadata without sacrificing consistency guarantees. 
A unified design also makes it possible to build a specialized engine, as we develop in this work, that can meet the specific requirements of a catalog in terms of both functionality and performance.

To this end, the proposed catalog itself can be seen as an independent database (sub-)system specialized for handling metadata operations. We characterize the primary use cases and present our prototype catalog engine, TreeCat, with the following design choices. (1) Given the hierarchical nature of metadata logical modeling and query access patterns, we adopt a {\em hierarchical data model} along with a {\em path navigation query language}. (2) For efficient query execution, we employ a storage layout inspired by file systems, implemented using {\em write-optimized key-value} stores, and leverage batch correlated scan operations. (3) We use versioned storage to support MVCC (multi-version concurrency control) for improved read performance, as well as version control operations such as time travel queries, clone, and snapshot. (4) We present a novel MVOCC mechanism that combines scan range locking and {\em precision locking} techniques. This approach efficiently manages predicate dependencies while minimizing validation costs. Additionally, we adopt a {\em commit-time update} technique to avoid aborts caused by conflicts on frequently updated fields, such as statistics. Although we focus on catalogs in this paper, some of our techniques are more generally applicable in any hierarchical database management system. TreeCat is implemented in about 12000 lines of C++ code, and exposes a gPRC interface to serve remote client requests. We present a comprehensive set of experiments illustrating the significant performance benefits of our new MVOCC protocol. We also show that TreeCat outperforms other state-of-the-art systems, including Delta Lake, HMS, and Iceberg. 

\begin{figure}[t]
  \centering
  \includegraphics[width=\linewidth]{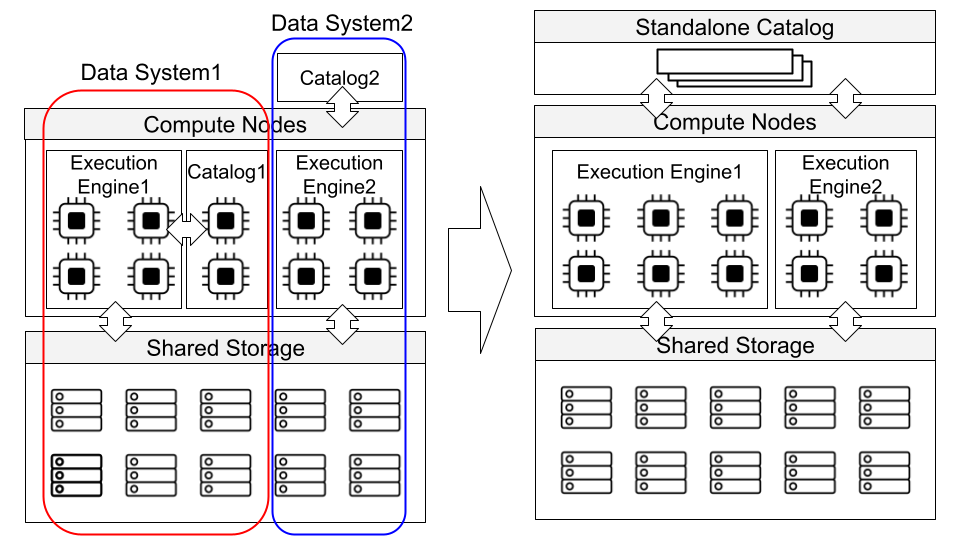}
  \caption{Disaggregating the catalog functionality for higher composability}
  \label{fig:catalogchange}

\end{figure}

\section{Background and Prior Work}
In this section, we categorize and classify existing catalog systems and highlight their shortcomings, which motivate the key requirements and features of TreeCat. The discussion of prior work related to more specific design and implementation details is interspersed in other relevant parts of the paper for better context.

We first distinguish between data catalogs and operational catalogs. {\bf Data catalogs} serve as an organization-level database of metadata about data assets across various systems. They can be viewed as an
    ``information retrieval system" for data assets. The main goal is to make the data assets of an organization more manageable and accessible, providing functions such as keyword search and retrieval of important contextual
        information. Data catalogs are typically implemented as federated systems that are updated asynchronously through data discovery processes or metadata ingestion
        pipelines~\cite{net_app,datahub_architecture,metacat_architecture,glue_architecture}. Because they do not directly serve database engines, they generally lack strict consistency or performance guarantees. There are
        numerous commercial data catalog software products, including those by Alation, Collibra, Informatica, AWS (Glue), Google (Dataplex), and Microsoft (Azure Data Catalog), as well as open source projects, such as Apache
        Atlas~\cite{atlas}, Metacat~\cite{metacat}, and DataHub~\cite{datahub}. {\bf Operational catalogs}, on the other hand, store metadata that are directly accessed by database engines during updates, query planning, and
        query execution. Although operational catalogs can be a part of, or serve as the main data source for, data catalogs, their primary functions are clearly different. In this paper, we focus exclusively on operational catalogs.

Catalogs can be 
either embedded or standalone. An {\bf embedded catalog} is integrated within a single data system, and external access to its metadata is restricted to the interfaces provided by the parent system. A {\bf standalone catalog},
on the other hand, exists as an independent service that is accessible by any compatible data system via a standard interface. 
Catalogs can be further differentiated based on their architectural designs. A catalog with a {\bf modular architecture} is composed of components spread across multiple autonomous services and systems. On the other hand, a catalog with a {\bf unified architecture} consists of tightly integrated components forming a single system. It is important to note that a unified system is not necessarily centralized; it can have a distributed architecture for scalability.

\begin{figure}[t]
  \centering
  \includegraphics[width=\linewidth]{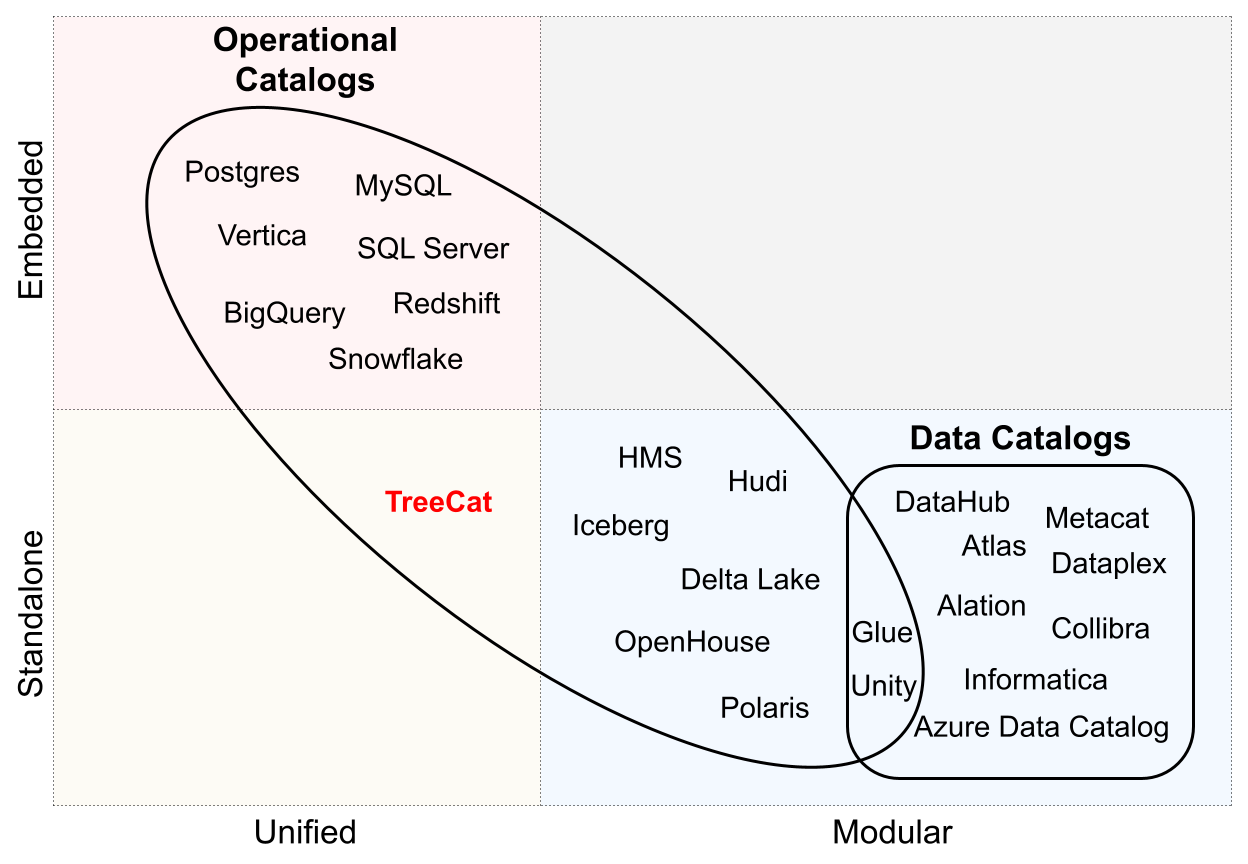}
  \caption{Classification of different catalog systems.}
  \label{fig:catalogclass}
\end{figure}

Traditional RDBMSs, including both transactional and analytical systems, have embedded catalogs with unified architectures. In
transactional RDBMSs, catalog metadata is stored as a predefined set of
relations~\cite{hellerstein2007architecture,postgresql_documentation,mysql_2016}. Apart from a specialized in-memory cache for low-latency reads, all
metadata use the same underlying storage representation and concurrency control mechanism as the data. Analytical RDBMSs and data warehouses have different implementations designed to handle increasingly decentralized scenarios. The catalog of a typical data warehouse functions as a specialized subsystem that decouples metadata from data, since the primary data processing infrastructure is not optimized for transactional workloads. For example, Snowflake, a cloud-native data warehouse, stores its metadata in FoundationDB~\cite{zhou2021foundationdb}, a distributed transactional key-value store. Many of its features, including automatic data clustering, data pruning during query execution, access control, and zero-copy clone, rely on catalog objects persisted in FoundationDB~\cite{snowflake_partition, motivala_2018}. Regardless of system performance, tight integration with the rest of the system makes it {\em difficult to extend the catalog functionality beyond a single system}~\cite{motivala_2018,armenatzoglou2022amazon,vandiver2018eon}.

Large-scale data processing engines~\cite{thusoo2010hive,armbrust2015spark,sethi2019presto} from the big data community use standalone catalogs with modular architectures. These systems originally emerged with a highly disaggregated architecture consisting of an independent compute engine, a distributed storage system, and a metadata service, each designed to be highly scalable. As part of the Hive ecosystem, HMS (Hive Metastore)~\cite{thusoo2010hive} was commonly used as the standalone catalog system. Backed by a transactional RDBMS (typically MySQL or PostgreSQL), HMS stores information about databases, tables, access control information, statistics, and partitions (groups of file objects organized by common attribute values). However, HMS does not store file-level metadata in the back-end RDBMS. Instead, it uses the shared storage system (distributed file system or object store) as an index on the files by grouping them in the same partition under the same directory (hence, it is modular). To scan a table, HMS must retrieve the paths of relevant partitions from the back-end RDBMS and call {\em list()} on each partition directory to obtain the list of files to scan. In an environment where a single dataset contains over a million files, the high latency of {\em list()} operation of cloud object stores like S3 becomes a serious performance bottleneck~\cite{armbrust2020delta}. In addition, even retrieving the list of a large number of partitions from HMS is slow due to an inefficient Thrift server implementation and suboptimal back-end database schema~\cite{katz_2021}. 

As the limitations of HMS became clear, lakehouse storage formats, namely Delta Lake, Hudi, and Iceberg~\cite{armbrust2020delta,apachehudi,apacheiceberg} emerged as alternatives. The specific implementations differ, but the common idea is to store metadata about file objects as separate {\bf metadata files} along with the data files in shared storage. By directly fetching and processing these metadata files, the performance bottlenecks of HMS can be avoided. Furthermore, additional file metadata, such as statistics, can be stored in metadata files and utilized for optimizations like file pruning, further improving system performance. Delta Lake uses a data structure called Delta log, which consists of a sequence of delta JSON files (with incrementing numerical IDs) and periodic checkpoint files~\cite{armbrust2020delta}. Iceberg organizes metadata into a hierarchy of files: a metadata file contains a list of manifest list files; a manifest list file contains a list of manifest files; and a manifest file contains a list of file objects~\cite{iceberg_spec}. Hudi identifies files by the timestamps at which they are first added and stores metadata in HFile format, enabling fast file pruning and listing~\cite{hudi_spec}.

Storing the metadata alongside the data has benefits such as a simplified architecture, good scalability due to the shared storage system, and speedup from parallel metadata processing. However, the notion that the adoption of the lakehouse storage format results in a simpler architecture than HMS is only partially true. A separate system must still be deployed to manage the metadata about high-level catalog objects, such as databases and tables; lakehouse storage formats simply offload most of the catalog functions to the shared storage system. More recently, several efforts have focused on providing these services through Hive-compatible APIs or REST APIs~\cite{iceberg_rest,unity,openhouse,polaris}. {\em Due to the modular architecture,  it is still quite difficult to maintain consistency of the metadata.} As discussed in Section~\ref{sec:concurrency control related work}, no lakehouse storage format currently supports multi-table transactions. Lastly, every metadata operation involves reading and writing metadata files in the shared storage system, which has non-trivial overhead. The high latency of these operations can limit the write throughput under high contention~\cite{jain2023analyzing}.

\section{High-Level Design}
\label{sec:design}
In this section, we present the high-level data model and the basic API of TreeCat. We also characterize the primary use cases with concrete examples.

\subsection{Data Model}

\begin{figure}[t]
  \centering
  \includegraphics[width=\linewidth]{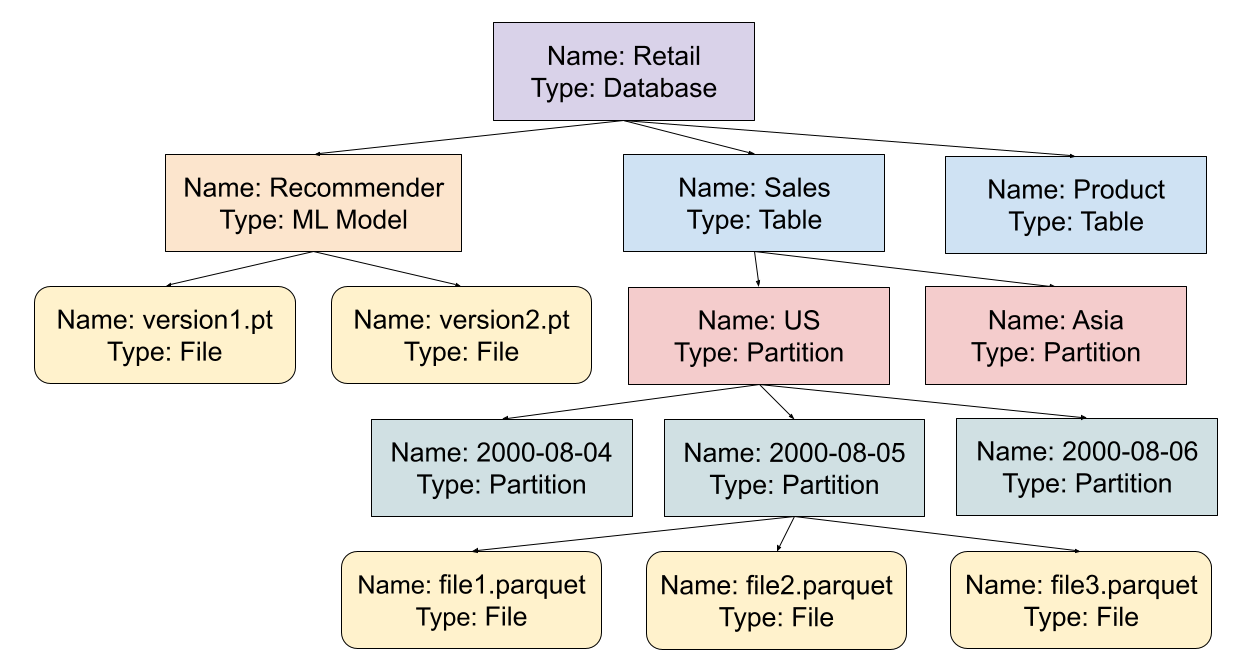}
  \caption{An example retail database, which contains heterogeneous assets, including relational data and ML models.}
  \label{fig:database}
\end{figure}

A catalog needs to manage metadata about numerous logical objects and their relationships, which are almost always organized into nested logical groups, naturally forming
    a multi-level hierarchy. It also needs to maintain arbitrary statistics and clustering information 
(for cost-based optimizations) and must incorporate explicit versioning semantics to support operations such as time-travel and cloning. Further, it should support a sufficiently powerful query language and/or API so that the clients don't end up replicating the functionality, while being flexible and extensible enough to support diverse data sets and database engines. This naturally leads us to adopt a {\em hierarchical data model} for TreeCat. We use an example Retail database shown in \autoref{fig:database} to illustrate this point.

\textit{Example 1: The retail database primarily contains relational data, but also includes other types of assets, such as ML models. Furthermore, different types of assets have different nested substructures. While an ML model consists of different versions, each identified by a tag, data in a relational table are partitioned by a set of attribute values. Leaf objects correspond to physical data objects in the storage system.}

We further note that: (1) Each group has common contextual information that is important for handling the constituent entities. (2) A predicate is often evaluated against such information, allowing entire groups of irrelevant objects to be pruned out during the data retrieval process. (3) There is data locality where entities in the same group are likely to be accessed together. These points are illustrated by the example use case that we present later. We now proceed with formal definitions of TreeCat's data model.
\begin{enumerate}[itemindent=0pt, left=0pt]
    \item Every node in the hierarchy corresponds to an object of a type with a set of properties.
    \item There are 2 types of objects:
    \begin{enumerate}[itemindent=0pt, left=0pt]
        \item A non-leaf object has a parent and a set of children objects. It is uniquely identified by its path, $\slash oid_1 \slash oid_2 \slash .. \slash oid_n$, where $oid_n$ is its object id and the object with the path of $\slash oid_1 \slash oid_2 \slash ... \slash oid_{n-1}$ is its parent.
        \item A leaf object has one or more parents, but \textbf{does not} have any child object. A leaf object with multiple parents has multiple paths. Given one of its paths, $\slash oid_1 \slash oid_2 \slash .. \slash oid_n$, $oid_n$ is one of its object ids and the object with the path of $\slash oid_1 \slash oid_2 \slash ... \slash oid_{n-1}$ is one of its parents. Unlike non-leaf objects, a leaf object is {\bf immutable} and can only be added or removed. 
    \end{enumerate}
    \item Every change (object addition, removal, or update) is associated with a $vid$ (version id) that determines its globally consistent chronological order.
\end{enumerate}


We highlight a few important design decisions that deviate from standard practices. 
(1) \textbf{Immutable leaf objects:} Leaf objects, which usually correspond to physical data objects (e.g., Parquet files), are treated as immutable to enable seamless versioning and sharing — updates can be handled as delete‐plus‐insert if needed.
(2) \textbf{Shared parents:} Allowing leaf objects to have multiple parents lets datasets be shared, cloned or branched, without duplicating physical files or metadata.
(3) \textbf{Global chronological versions:} A single, consistent timeline underlies our concurrency control and powers version‐control features such as time‐travel queries, snapshots, and clones.

Note that, while Figure~\ref{fig:database} shows one possible instantiation of metadata schema, the engine imposes no fixed metadata model—users choose whatever schema best fits their use case.

\subsection{Query Language} \label{subsec:query}
TreeCat's query language is based on top-down path navigation, where a query takes the form of a path expression, a sequence of $n$ predicates, each of which is evaluated against the children of context objects that satisfy the preceding path subexpression. A predicate is either a wildcard character or an expression (enclosed with $[]$) that may be evaluated against the object id or object properties. We extend the example in \autoref{fig:database} to demonstrate how client systems can utilize TreeCat. Consider how a relational database system would execute the following simple query.

\begin{verbatim}
SELECT count(*)
FROM retail.sales
WHERE region = 'US' AND date > '2025-01-01' 
AND price > 5;
\end{verbatim}

\noindent After parsing the query, the system has to semantically validate it. First, it retrieves the metadata of the Sales table from TreeCat, using the following metadata query.

\begin{verbatim}
/[obj_id='retail']/[obj_id='sales' 
    and obj_type='table']
\end{verbatim}

\noindent A nonempty query result also validates that the table with the specified database name and table name indeed exists. After the table metadata is retrieved, the system uses it to further check whether column names and types are valid, etc. During the query planning phase, the paths and statistics of the data files that must be scanned during the query execution have to be retrieved. The following query is submitted to retrieve the file-level metadata that satisfy the appropriate predicate filters.

\begin{verbatim}
/[obj_id='retail']/[obj_id='sales' and 
    obj_type='table']/[part_val='US']
    /[part_val > '2025-01-01']/[stats.price.min > 5]
\end{verbatim}

\noindent The physical plan can finally be executed, once it is constructed.

We briefly discuss related work, including the APIs of existing standalone catalogs and query languages of similar, more general-purpose database systems. Standalone catalogs, such as HMS and lakehouse storage formats, provide basic RPC routines, such as {\em getTable()} and {\em getPartitions()}, via Thrift or the REST API for retrieving high-level catalog objects. Retrieving file-level metadata of a given table involves iterating the list of file paths via an iterator interface with an optional predicate filter, rather than a unified query language like TreeCat. Meanwhile, the query languages of general graph databases, including Cypher~\cite{cypher}, Gremlin~\cite{gremlin}, and GSQL~\cite{gsql}, and those of XML databases, including XPath~\cite{clark1999xml} and XQuery~\cite{xquery}, are more powerful than TreeCat's query language and can express a wider range of path queries. However, advanced semantics makes efficient implementation of a strong transaction isolation level very difficult. To our knowledge, there are currently no efficient graph database or XML database that fully support serializable isolation level. In our design, we sacrifice the query expressivity, so we can guarantee serializability using techniques that are discussed in Section~\ref{sec:concurrency}.
 
\subsection{System API} \label{subsec:api}
\noindent\ul{\bf Basic API:} Next, we list a basic API that TreeCat supports. 
\begin{itemize}[itemindent=0pt, left=0pt]
    \item \texttt{startTransaction(mode)}: Starts a new transaction in the specified mode. If the mode is read-only, it returns the latest $vid$ at which a snapshot of the catalog can be read. If the mode is read-write, it creates a new transaction session and returns $transaction\_id$ and $read\_vid$.
    \item \texttt{executeQuery(query, optional transaction\_id, optional vid)}: Executes a read query on the catalog using the query language defined above.
    \item \texttt{commit(transaction\_id, write\_set)}: Validates and commits the transaction identified by $transaction\_id$, applying modifications specified by the given $write\_set$. 
    \item \texttt{snapshot(snapshot\_name, vid)}: Creates an immutable snapshot of the catalog data at $vid$ and names it $snapshot\_name$.
    \item \texttt{clone(src\_path, dest\_path, optional vid)}: Clones/copies the object at $src\_path$ and all of its descendants to the $dest\_path$. Note that this is a metadata-level operation, and does not involve cloning the physical objects.
\end{itemize}

\smallskip

\noindent\ul{\bf Updates:} TreeCat can be updated via {\em commit()} function, which takes the modifications in the form of a write set. A write set is a collection of \texttt{(path, value, type)} tuples, where \texttt{path} specifies the object to modify, \texttt{value} contains the updated value, and \texttt{type} specifies the operation type, as listed below:

\begin{enumerate}[itemindent=0pt, left=0pt]
    \item \textbf{Add:} a new object of the given path and initialize it with the given value. The operation has a precondition that the parent object exists, but an object with the same path does not. 
    \item \textbf{Update:} the object with the given path with the given value. A new object is created if the path does not exist.
    \item \textbf{Remove:} the object with the given path and all its descendants. 
    \item \textbf{Merge:} Also known as the commit-time update operation~\cite{huang2020opportunities}, merge operation turns a read-modify-write operation into a single blind write operation that is performed at commit time. The given object value is a delta, rather than the new object value, and is applied to the existing object value at commit time. The operation has a precondition that the object already exists. 
\end{enumerate}

\smallskip

\noindent We describe the merge operation in more detail and highlight its importance. Merge operation is useful when objects are frequently updated by commutative read-modify-write operations (e.g., counters), which may result in many conflicts. An important use case in TreeCat is applying changes in statistics that propagate up the hierarchy, which can result in contention on high-level statistics objects. We present an example of an update to table statistics that follows data insert:

\smallskip

\noindent\textbf{Original Value:} \texttt{\{...size:1487..min:3...\}} 

\noindent\textbf{Delta:} \texttt{\{...size:\{op:+,val:124\}..min:\{op:min,val:0\}...\}}

\noindent\textbf{Final Value:} \texttt{\{...size:1611..min:0...\}} 

\smallskip

\noindent Currently, the types of delta operations TreeCat supports include addition, subtraction, min, and max for a limited number of types.

\section{System Architecture}
\label{sec:architecture}

We first outline the high level architecture of TreeCat, then elaborate on the core system components. We discuss the concurrency control mechanism in Section~\ref{sec:concurrency} in greater detail. 

\subsection{High Level Architecture}
TreeCat is designed as a single-server database engine that handles incoming requests from remote client systems via the gRPC interface. Key components include: 

\begin{enumerate}[itemindent=0pt, left=0pt]
    \item \textbf{Backend Server:} Provides the API described in the previous section and schedules client tasks to the main thread pool.
    \item \textbf{Executor:} Contains utilities for creating the execution plan of a read query and executing it. 
    \item \textbf{Transaction Manager:} Manages the overall concurrency mechanism. It keeps track of ongoing transaction states and launches a validation and commit process of the transaction upon receiving a commit request.  
    \item \textbf{Storage:} All catalog data is stored in the storage layer, which provides APIs to get an object by path and traverse all children of an object. The underlying storage engine is RocksDB~\cite{rocksdb}. 
\end{enumerate}

This architecture allows a single TreeCat instance (possibly replicated, see Section~\ref{sec:discussion}) to handle the metadata for several
different logical databases, and concurrently serve many execution engines while ensuring metadata consistency. Multiple TreeCat instances could be spun to serve different
logical databases for administrative autonomy; however, any logical database (i.e., a collection of logically related tables or datasets) should
ideally be served by a single TreeCat instance so that cross-table consistency can be enforced.

\begin{figure}[t]
  \centering
  \includegraphics[width=\linewidth]{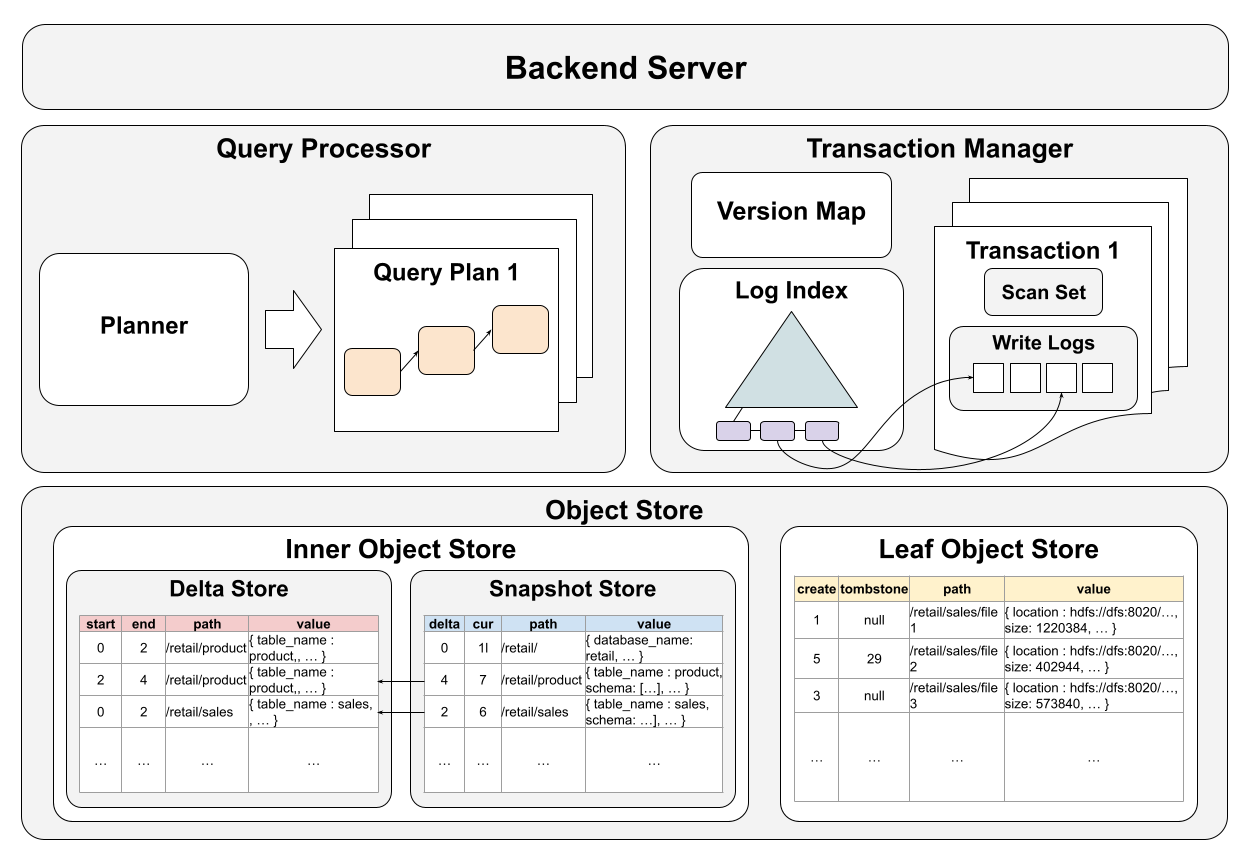}
  \caption{High-level architecture of TreeCat}
  \label{fig:architecture}
\end{figure}

\subsection{Backend Server}
We use gRPC~\cite{grpc}, a Remote Procedure Call (RPC) framework developed by Google, to provide the APIs listed in the previous section. gRPC is language-agnostic and offers portability across different languages. While REST API is another option with good portability, gRPC provides better performance due to its use of pre-compiled serialization and deserialization routines and support for server-side streaming, especially for large data loads (it is not uncommon for reads in this context to return MBs of data).

\subsection{Storage Engine}
When selecting the underlying storage engine, we considered several key requirements, including (1) optimizations for efficient traversal of hierarchical data, (2) efficient implementation of explicit versioning semantics, (3) fine-grained concurrency control with strong guarantees that can efficiently handle frequent updates to hierarchical data, and (4) embedded storage for low latency.

Numerous options exist for full-fledged database systems, ranging from RDBMS with XML~\cite{shanmugasundaram2001general,tatarinov2002storing,o2004ordpaths} and graph~\cite{ten2023duckpgq} extensions, to graph databases~\cite{neo4j,nebulagraph,hypergraphdb,arangodb,janusgraph,dubey2016weaver,zhu13livegraph,de2021teseo,fuchs2022sortledton}, and document stores~\cite{mongodb,couchdb,arangodb}. However, meeting all the requirements, especially (2) explicit versioning and (3) fine-grained concurrency control, proves to be difficult, as these systems are designed and implemented for more general use cases. As discussed before, lack of efficient support for serializable isolation level is especially problematic for graph databases and document stores.

After careful consideration, we built TreeCat on top of RocksDB, a write-optimized key-value store. 
Inspired by file systems built on top of write-optimized key-value stores~\cite{jannen2015betrfs,shetty2013building,esmet2012tokufs}, we adopt a storage layout where objects are sorted by the full object path, first by the path depth, then lexicographically. This layout allows fast listing of sibling objects by performing a range scan on the common prefix. It has advantages over storing the parent-child relation as a separate set of edges, which incurs the overhead of joining the objects (vertices) with edges during traversals. 
Although each range scan incurs a seek cost, the expense is amortized since objects typically have multiple children; as we show later, query cost is bounded by the query's selectivity because most irrelevant objects are pruned early.
This layout is also more suited for versioning and transactional updates compared to the scheme where an object's children are stored together as a single list. With the latter, installing a new object version forces installation of an updated version of the parent's children list, potentially cascading changes to the root and causing read–write conflicts that severely limit concurrency under serializable isolation.
{\em This also imposes a fundamental limitation of Iceberg for supporting multi-table transactions.}

\begin{figure}[t]
  \centering
  \includegraphics[width=\linewidth]{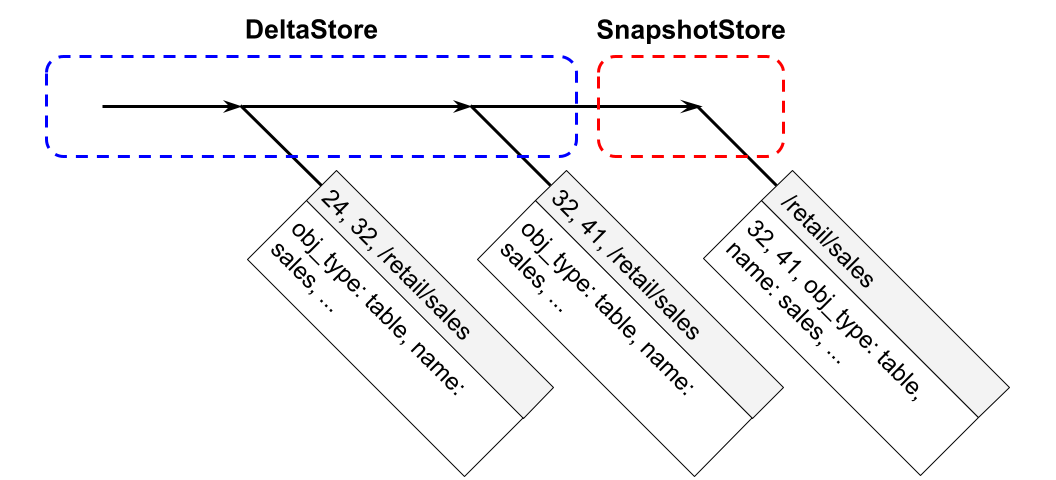}
  \caption{Example storage layout of the Sales table from ~\autoref{fig:database} in {\bf InnerObjectStore}.}
  \label{fig:storagelayer}
\end{figure}

Object values are stored in BSON format~\cite{bson}, a binary JSON format that originated from MongoDB, is queriable, and can easily be converted to JSON format. While other efficient JSON-compatible binary formats exist (e.g., PostgreSQL’s JSONB), we chose BSON for its robust library support and rich feature set, which enables faster system development. We use two separate storage systems, {\bf InnerObjectStore} and {\bf LeafObjectStore}, which store non-leaf objects and leaf objects respectively. {\bf LeafObjectStore} stores one object value for each path and the $vid$ range during which the object is visible. Unlike leaf objects, multiple past versions of non-leaf objects are persisted. For non-leaf objects, we adopt the ``time-travel" storage scheme~\cite{wu2017empirical} where the most recent version is stored in the main {\bf SnapshotStore} and the chain of past versions are stored in a separate {\bf DeltaStore}, as illustrated in \autoref{fig:storagelayer}. {\bf SnapshotStore} has the following mapping:

\smallskip

\begin{center}
path $\rightarrow$ delta\_vid, cur\_vid, object value  
\end{center}

\smallskip

\noindent where $cur\_vid$ is the $vid$ from which the current snapshot is valid, and $delta\_vid$ is the $start\_vid$ of the previous object value. {\bf DeltaStore} has the following mapping:

\smallskip

\begin{center}
start\_vid, end\_vid, path $\rightarrow$ object value  
\end{center}

\smallskip

\noindent where the object value is visible between $start\_vid$ and $end\_vid$. In {\bf DeltaStore}, the objects are sorted by the object path, then by $start\_vid$ in descending order, so past object values can be traversed through a range scan. Retrieving the object value of a non-leaf object involves first checking the {\bf SnapshotStore} for the current value and traversing the past values in {\bf DeltaStore}, if necessary. Retrieving past versions incurs the cost of version traversal, but we expect most operations will access the most recent version. {\bf LeafObjectStore} has the following mapping: 

\smallskip

\begin{center}
path $\rightarrow$ create\_vid, tombstone\_vid, object value/primary path  
\end{center}

\smallskip

\noindent where the object is visible between $create\_vid$ and $tombstone\_vid$. 
Because leaf objects can have multiple parents, there could be a layer of indirection where the path could map to the primary path, which, in turn, maps to the object value.

\subsection{Query Execution}
A read query is executed as a sequence of physical operators analogous to the correlated join operator from RDBMSs (also called CROSS APPLY or LATERAL JOIN), which passes each value from the outer query to a user-defined function or a subquery and evaluates it. Unlike other standard join operators, the correlated join operator allows for early pruning of objects in the inner query by only scanning objects that join with objects that satisfy the outer query. We use this mechanism to avoid scanning large number of leaf objects whose ancestors do not satisfy the preceding predicates in the path expression. For example, while executing \texttt{/[obj\_id = `retail']/[obj\_id = `sales']/*/*/*}, file objects that belong to the Customer table need not be scanned. 

\begin{figure}[t]
  \centering
  \includegraphics[width=\linewidth]{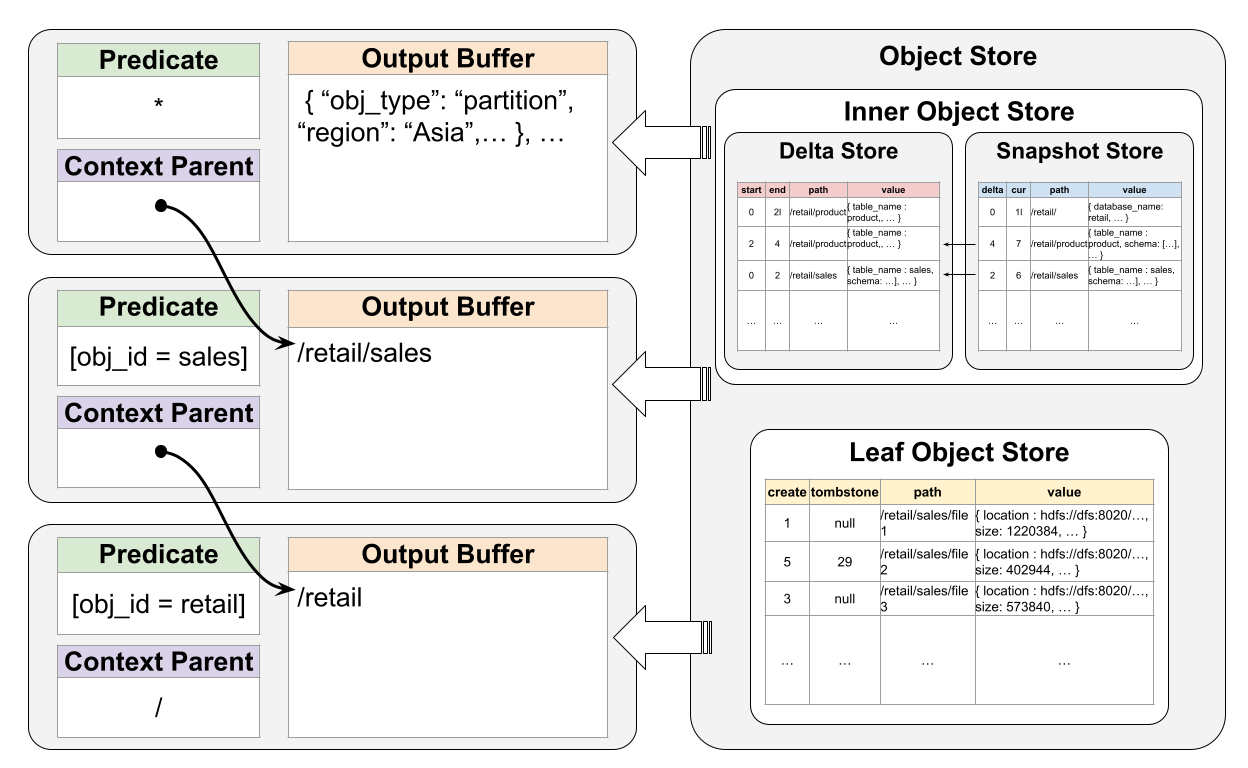}
  \caption{Query Execution of a Simple Example Query: \texttt{/[obj\_id = retail]/[obj\_id = sales]/*}. The output buffer in the top most node holds the final result set.} 
  \label{fig:execution}
\end{figure}

Given a path expression of depth $n$, the execution plan is constructed as a single chain of $n$ execution nodes, each with the corresponding predicate, a fixed-sized output buffer, a pointer to its child node, and handles to the object store. The query plan follows a standard batch iterator model where each node invokes $next()$ on its child node, retrieving a batch of context object paths. For each context object path, the executor scans and evaluates the predicate against every child object visible in the $read\_vid$. If the object satisfies the predicate, its path (or its value if the last execution node) is appended to the output buffer. The execution of a simple example query is illustrated in \autoref{fig:execution}. The planner currently applies one optimization rule where the lower and/or upper bounds of the object store iterator are set if the predicate is only satisfied by a range of object ids (e.g. \texttt{[1 < obj\_id < 100]}).

Assuming that the underlying data takes the form of a balanced tree $T$ with a constant fan-out $f$ and height $h$, we can compute the cost of a path query $q$ of depth $d$ where $selectivity(q) = (s_1, s_2,.., s_d)$ specifies the list of predicate selectivities. We also assume independent selectivities for simplicity. The two main cost components are the scan cost and seek cost of every range scan. We first compute the total number of scan operations. At $i^{th}$ level of the query evaluation, the number of scanned objects increases by a factor of $s_{i+1}f$:

\begin{align*}
    n_{scan} &= f + s_1f^2 +...+ \prod_{i=1}^{d-1}s_{i}f^{d} \leq \frac{s^df^{d+1} }{sf - 1} 
    \leq s^df^{d+1}
\end{align*}


\noindent where $s = \max_{i=1}^{d-1} s_i$ and $sf\geq2$. The number of seek operations is equal to the number of scan operations of the preceding path sub-expression and can be computed to be $s^{d-1}f^d$ in a similar manner. Because there is no closed form formula for $f$ in terms of $|T|$, we approximate $f$ to $|T|^{1/h}$. Finally, the total cost can be approximated to $s^{d-1} \cdot |T|^{d/h}(s \cdot |T|^{1/h} \cdot c_{scan} + c_{seek})$ where $c_{scan}$ and $c_{seek}$ represent the cost of a single scan operation and that of a seek operation, respectively. We observe that the multiplier effect of selectivity $s$ effectively bounds the number of scan and seek operations necessary for executing the given query, which aligns with the intuition about ``early pruning" of objects. This contrasts with more popular join operators, such as sort-merge join or hash join that require at least one scan of the entire table, which can be costly for a large table size (analogously tree size).

\subsection{Version Control Operations}
The system supports globally consistent chronological versioning of the entire catalog data, using monotone increasing $vid$ associated with every transaction. $snapshot()$ operation is implemented by maintaining an additional mapping between the snapshot name and the $vid$, which can be used to retrieve the correct version. $clone()$ operation copies the source object and all of its descendants (except for the leaf objects) at some $vid$ to the destination path. For cloned leaf objects, the system adds a mapping from the new destination paths to the primary paths, so they are shared between the source and destination parents, saving the storage costs of duplication.

\section{Concurrency Control}
\label{sec:concurrency}
In this section, we present TreeCat's concurrency control mechanism, which guarantees serializability. Besides our use case in TreeCat, the scheme is also of independent interest in a more general context of databases with a hierarchical data model. 

\subsection{Discussion}
We first discuss the background and primary motivations, followed by an in-depth description of the implementation details. We adopt the MVOCC scheme where read-only transactions operate on a versioned snapshot of the data, and read-write transactions follow a protocol similar to the standard OCC mechanism. Because we implemented MVOCC on top of the existing version system, $vid$ is synonymous with $timestamp$ as commonly discussed in the concurrency control literature. For read-only transactions, we follow the common practice of assigning a $read\_vid$, the version up to which data are safe for read, at the start of the transaction, and executing queries against the versions of objects visible at the assigned $read\_vid$. Since this aspect of the mechanism aligns with other existing systems, our main focus is on the OCC, which utilizes a predicate-based method, often overlooked by standard schemes.

The important feature of TreeCat is that read queries involve {\bf predicate read} operations. Unlike simple key-value read and write operations, predicate read operations can introduce {\em predicate
dependencies}~\cite{adya2000generalized,clark2024validating}, which are distinguished from the {\em item dependencies} (write-read, write-write, and read-write) that are commonly discussed in the concurrency control
literature. The infamous {\em phantom read anomaly} is one example of a predicate anomaly that can occur if predicate dependencies are not handled correctly by the concurrency control
mechanism~\cite{eswaran1976notions}. 
 Unlike item dependencies—which are based on conflicting operations on the same item—predicate dependencies occur when an update alters whether an object satisfies a given predicate.
Specifically, if a write operation's before-image satisfies the predicate of a predicate read operation but its after-image does not, or vice versa, a predicate dependency exists between the two operations~\cite{adya2000generalized}. Item-based approaches, including record locks in 2PL (2-Phase Locking), read and write sets in OCC, and time stamps on individual items in TO (timestamp ordering), lack semantics for predicate reads, and cannot handle predicate dependencies without additional mechanisms. 

The most common way to augment these schemes to handle predicate dependencies is to lock the entire scan range of the predicate read operation. If an index scan is used, the scan range can be limited to a range of values of the indexed attribute on which the predicate is defined.  However, this optimization cannot be applied to arbitrary predicates, which may be a complex expression or involve attributes on which an index does not exist. The scheme is, at best, a coarse-grained approximation that may lock an irrelevant range of values, resulting in unnecessary conflicts. 

We illustrate this point with a simple example. Let $R$ be a logical {\em table} that is partitioned into multiple {\em files}. Transaction $T_1$ performs a predicate read operation on $R$'s file metadata with the predicate $p$ on the file statistics. Because there is no index on the file statistics, $T_1$ scans the entire file metadata of $R$. Meanwhile, $T_2$ inserts a file $f$ with file metadata that does not satisfy $p$ into $R$. There is no reason for the two operations to conflict, because the new file $f$ is never visible to $T_1$'s predicate read, regardless of the order of operations. However, with scan range locking, $T_1$ would lock the entire range of $R$'s file metadata. In a pessimistic scheme (S2PL), either of the two transactions would block the other. In an optimistic scheme (OCC), if $T_2$ is submitted for validation before $T_1$, $T_1$ would be aborted. 

There are predicate-based methods that can precisely detect any predicate dependency. One scheme is to re-execute the scan operations to observe any changes in the visible objects~\cite{larson2011high}. While precise, re-executing scans can be costly for large ranges. Precision locking~\cite{jordan1981precision} is a form of predicate locking that evaluates the predicate directly against posted write operations. This technique detects predicate dependencies accurately and avoids the satisfiability test problem of conventional predicate locking. But in TreeCat, a read query is correlated whence an object satisfies the entire path expression only if all of its ancestors also satisfy the corresponding path sub-expressions. A naive implementation that only considers the predicates in a given query can, in fact, result in false conflicts.  

In our scheme, we combine scan range locking with precision locking, leveraging the strengths of both methods. We first apply optimistic prefix-based scan range locking to detect candidate conflicts, i.e., write operations that fall within the scan range, but may not necessarily cause conflicts. Then, we use optimistic precision locking to confirm whether the candidate is actually a conflict. This scheme achieves the granularity of precision locking while improving efficiency by pruning unnecessary predicate evaluations with scan range locking. In addition, we make use of an in-memory index structure to quickly locate only the log records that fall in the scan range, rather than applying precision locking on the entire set.

\subsection{Implementation}
The concurrency control is handled by the \textbf{TransactionManager}, which keeps track of $read\_vid$, the latest version up to which, the catalog can be read safely, and $commit\_vid$, which is the monotone increasing vid assigned to a transaction before final commit. Key in-memory data structures include:   

\begin{enumerate}[itemindent=0pt, left=0pt]
    \item $ScanSet$: Maintained by each transaction. Stores pairs of the scan range (in the form of the context object path) and the predicate of the predicate read operations executed as part of the transaction.
    \item $LogBuffer$: Maintained by each transaction. Stores before- and after-images of all recent write operations.
    \item $VersionMap$: Used by \textbf{TransactionManager} for validation. Maps object path to the most recent $vid$ of its children.
    \item $LogIndex$: Used by \textbf{TransactionManager} for validation. Index on the $LogBuffer$, mapping pairs of object path and $vid$ to the log records. The index is implemented using a lock-free skip list and is sorted by the parent path, then $vid$.
    
\end{enumerate}

When $startTransaction()$ is called, a new transaction session is created and $read\_vid$ is loaded as the vid at which all read queries are executed in the session. During query execution, pairs of predicates and parent paths (prefix) of objects scanned for predicate evaluation are added to the $ScanSet$. Also, a private copy of the write set is constructed on the client side. When the transaction is submitted for the commit process, the write set is sent to the server and undergoes preprocessing, where it is copied to the $LogBuffer$ and preconditions are checked. If the transaction passes these checks, the transaction is scheduled for the main commit process.

The main commit process consists of two phases: validation and batch write. For each phase, there is a dedicated thread pool of a fixed size, with each thread handling a single hash partition of the $ScanSet$ and the write set. At the start of the validation, \textbf{TransactionManager} executes a $fetch\_and\_add()$ operation on the $commit\_vid$ and assigns it to the transaction. Then, each worker thread validates the scan ranges in the $ScanSet$ against the $VersionMap$ to determine whether any objects scanned by predicate read operations have changed. If potential conflicts are detected, an iterator on $LogIndex$ is set to scan and apply precision locks on corresponding recent writes with $vid$ greater than the transaction's $read\_vid$ and fall in the scan range. If any conflict is detected through precision locking, the transaction is aborted. After all threads pass validation, they update the $VersionMap$ and $LogIndex$ based on their respective write set partitions for the transaction. The validation phase (omitting the initial $vid$ assignment and latter updates to in-memory data structures) proceeds as per Algorithm~\ref{alg:validation}. Note that for merge operation, before- and after-image may have to be constructed as they are not available until the final commit. However, once constructed, the images are cached so they are accessible during the validation of subsequent transactions.  

Once validation is complete, the transaction proceeds to the write phase. During this phase, each thread processes its assigned write partition, evaluating the final values of merge operations (commit-time updates) and applying the writes as an atomic batch write operation to the underlying RocksDB instance. After every thread finishes its batch write, WAL records of the transaction are flushed to disk along with those of other transactions via $SyncWAL()$ operation, which calls $fsync()$ on the WAL files. The $read\_vid$ is also updated, making the committed changes visible for subsequent reads. Transactions are processed one at a time in each phase, but parallelism is achieved via hash-partitioning the workload and pipelining different phases of the commit process. 

Old $LogBuffer$s and entries in $VersionMap$ and $LogIndex$ that are no longer needed are asynchronously garbage collected by a background thread. For this purpose, the \textbf{TransactionManager} keeps track of $watermark$, the oldest $read\_vid$ among the current transactions. After each transaction commit, the $watermark$ is updated and all entries in the aforementioned data structures with $vid$ below the $watermark$ are garbage collected. 

\begin{algorithm}[t]
\caption{Validation Algorithm}
\label{alg:validation}
\begin{algorithmic}[1]
    \STATE \textbf{for} $(path,predicate) \in txn.ScanSet$
        \STATE \hspace{2em} // scan range locking
        \STATE \hspace{2em} \textbf{if} $VersionMap[path] > txn.read\_vid$
            \STATE \hspace{4em} $iter \gets LogIndex.newIterator()$
            \STATE \hspace{4em} $iter.seek(path, txn.read\_vid)$
            \STATE \hspace{4em} \textbf{while} $iter.path.parent = path$
            \STATE \hspace{6em} \textbf{if} $iter.type = merge$ and $iter.image.empty()$
                \STATE \hspace{8em} $iter.image.construct()$
            \STATE \hspace{6em} // precision locking
            \STATE \hspace{6em} \textbf{if} $iter.image.before \models predicate$
                \STATE \hspace{8em} $abort(txn)$
            \STATE \hspace{6em} \textbf{else if} $iter.image.after \models predicate$
                \STATE \hspace{8em} $abort(txn)$
\end{algorithmic}
\end{algorithm}

\subsection{Proof of Correctness}
In this section, we prove that TreeCat's concurrency control mechanism guarantees serializable isolation.

\smallskip

\noindent\ul{\bf Concurrency Control Model:} We first formalize the concurrency control model. We follow Adya's model~\cite{adya2000generalized}, where every object in the database has one or more versions. Every operation on an object is mapped to exactly one version of that object. Given an object $x$, $x_i$ is a version with $vid$ $i$, installed by transaction $T_i$. Every object is assumed to exist forever regardless of when they are inserted or deleted. The unborn state is represented as version $x_{init}$, while the dead state is represented as $x_i$, installed by the delete operation of $T_i$. Every object is uniquely identified by its full path. 

It is important to discuss how TreeCat's query evaluation is represented in this model. Evaluation of a path expression involves performing a series of predicate range scans, each with respect to a context parent object. For example, given the path expression, \texttt{/[obj\_id = ``retail"]/[name = ``sales"]}, the evaluation of the second predicate, $name = ``sales"$, is executed as a range scan on the children of \texttt{/retail}. Each range scan can be modeled as a predicate read operation, where the predicate is a conjunction of the original predicate and the equivalence between the object parent path and the context parent path. Following the above example, evaluating $name = ``sales"$ against \texttt{/retail}'s children objects can be modeled as a predicate read operation with the predicate, $name = ``sales" \wedge parent = /retail$. The evaluation of a single predicate in a path query can correspond to multiple such predicate read operations, as there can be multiple context parent objects with which it is evaluated. 

\smallskip

\noindent\ul{\bf Main Proof:} We now proceed with the main proof.

\smallskip

\noindent{\bf Lemma 1.} \textit{For any two committed transactions $T_i$ and $T_j$ where $commit\_vid_i < commit\_vid_j$, there is no dependency from $T_j$ to $T_i$.}

\begin{proof}
We prove the lemma by enumeration. First, there cannot be any ww-dependency from $T_j$ to $T_i$ because write operations are performed in the order of the $commit\_vid$ in the write phase. Second, wr-dependency from $T_j$ to $T_i$ is also not possible. Given that $read\_vid_i < commit\_vid_i$ and $commit\_vid_i < commit\_vid_j$, $read\_vid_i < commit\_vid_j$ by transitivity. Because changes committed by $T_j$ become visible only after the $read\_vid$ has been incremented to $commit\_vid_j$, $T_i$, which has a lower $read\_vid$, cannot observe any changes installed by $T_j$. Lastly, we show by contradiction that there cannot be any rw-antidependency from $T_j$ to $T_i$. Assume that there is a rw-antidependency from $T_j$ to $T_i$. Consider the state when $T_i$ passed validation, but $T_j$ is about to enter validation. There must be a write operation $w$ executed by $T_i$ on an object $o$, whose before- or after-image satisfies the predicate of a predicate read operation $P$ executed by $T_j$, which scans $w$'s before-image. Per our earlier discussion, the predicate of $P$ is $p \wedge parent = c$, where $p$ is the predicate in the path query and $c$ is the context parent object. During the query execution of $T_j$, $(c,p)$ was added to $T_j$'s $ScanSet$. Also, $(o.parent, i)$ was added to $VersionMap$ and a mapping from $o$ to $w$'s before- and after-images was added to $LogIndex$ during the validation of $T_i$. During the validation of $T_j$, $c$ is found in $VersionMap$ because $c = o.parent$ holds true. Since $read\_vid_j < commit\_vid_i$, $p$ is evaluated against $o$'s entries in $LogIndex$. Because the before- or after-image of $w$ satisfies $p$, the rw-conflict is detected and $T_j$ is aborted, leading to a contradiction. We have now considered all types of dependencies, so the lemma holds true.
\end{proof}

\noindent{\bf Theorem 1.} \textit{Any schedule of committed transactions output by TreeCat is conflict serializable.}

\begin{proof}
We prove the theorem by contradiction. Assume that there is a schedule $s$, output by TreeCat, that is not conflict serializable. Then, the corresponding precedence graph, $G(s) = (V,E)$ must contain at least one cycle. Consider a directed edge $(T_i,T_j)$ in the cycle. Because there is a dependency from $T_i$ to $T_j$, $commit\_vid_i < commit\_vid_j$ by Lemma 1. Because $(T_i,T_j)$ is in a cycle there is a path from $T_j$ to $T_i$. Using Lemma 1 and transitivity, we can prove by induction (which we omit for brevity) that $commit\_vid_i > commit\_vid_j$, resulting in a contradiction.
\end{proof}

\begin{figure*}[ht]
  \centering
  \includegraphics[width=\linewidth]{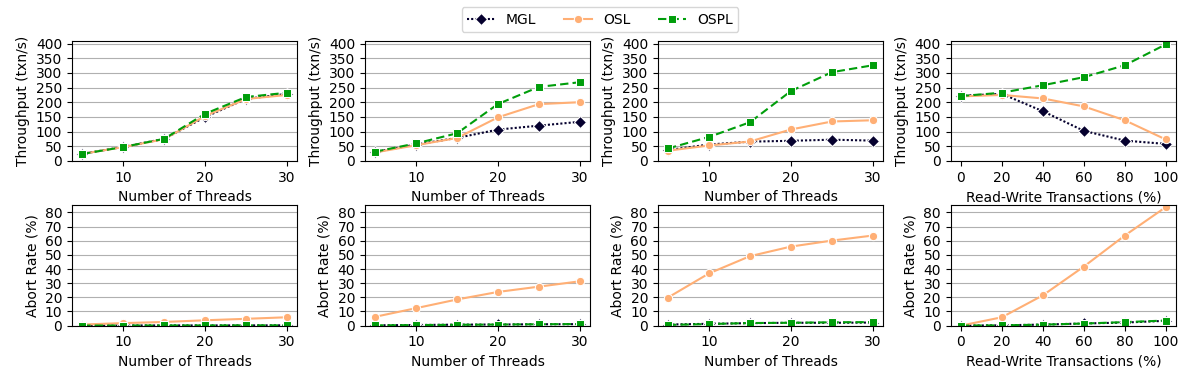}
  \subfloat[ Read-Intensive (80-20) ]{\hspace{.25\linewidth}}
\subfloat[Balanced (50-50)]{\hspace{.25\linewidth}}
\subfloat[Write-Intensive (20-80)]{\hspace{.25\linewidth}}
\subfloat[30 Threads]{\hspace{.25\linewidth}}
  \caption{Throughput and abort rate across different number of client threads and different types of workloads.}
  \label{fig:concurrency1}
\end{figure*}

\subsection{Related Work}
\label{sec:concurrency control related work}
We first discuss the concurrency control mechanisms of other catalog systems and then examine how predicate reads are handled by concurrency control mechanisms across more general data systems.  

\smallskip

\noindent\ul{\bf Catalog Concurrency Control:} Traditional RDBMSs store the catalog as a set of relations, relying on the same concurrency mechanism that is used for data. While data warehouses use proprietary mechanisms, Snowflake is known to use FoundationDB to persist transaction states and lock queues~\cite{motivala_2018}. Even though resource locks at the level of micro-partitions are available, most single table DML queries require locking all micro-partitions~\cite{snowflake_locking}. Delta Lake uses snapshot-based OCC where, during the final commit, it tries to atomically write the next delta log record. If the underlying storage system does not support atomic file rename, it has to resort to another transactional system, such as DynamoDB~\cite{sandre_lee_kryński_2022}. Delta Lake currently only supports single-table transactions. However, there is early work on introducing a new commit protocol called {\em{commit coordinator}}, an external system that is dedicated to handling concurrency control, which may support multi-table transactions in the future~\cite{das_portis_sandre}. Iceberg also uses snapshot-based OCC where, after writing all the metadata files (manifest files, manifest list files, and metadata file), it does a final commit by switching the current pointer in the catalog to the new metadata file. If there is a conflict, the transaction undergoes a retry validation, so the entire operation need not be repeated from scratch~\cite{okolnychyi2024petabyte}. Like Delta Lake, Iceberg's commit protocol only supports single-table transactions. Recently, multi-table transactions have been added to the Iceberg REST API. But as far as we know, Nessie~\cite{nessie}, which supports git-like versioning, is the only compatible catalog that supports it. Even so, the transactions operate at the level of a single branch of the whole database. Lastly, Hudi also uses snapshot-based OCC and does conflict resolution based on the files modified by multiple writers~\cite{hudi_concurrency}. Hudi requires an external lock provider for locking tables and can only support single-table transactions. 

\smallskip

\noindent\ul{\bf Predicate Read:} To guarantee serializable isolation, a data system must account for any predicate dependencies~\cite{adya2000generalized}. The standard solution is to lock the entire scan range of the predicate read operation, an example of which is next key locking~\cite{mohan1989aries}, which locks the current record and the gap until the next record. Since taking many record locks can cause excessive overheads, multiple granularity locking~\cite{gray1976granularity} is commonly used to escalate locks to more coarse-grained ones, such as table locks. Large scan range may cause unnecessary conflicts, so it is crucial to minimize the scan range. For simple predicates, performing an index scan can limit the scan range. A more advanced technique may be used to better approximate the minimum scan range for complex predicates. For example, Spanner uses a runtime data structure called a filter tree to compute the approximate scan range~\cite{bacon2017spanner}. One way to precisely handle predicate dependencies is to validate the scan operation by re-executing it, which is adopted by Hekaton~\cite{diaconu2013hekaton}. Another approach is to use predicate-based locking. Conventional predicate locking involves satisfiability test problem, which is known to be NP-complete~\cite{hunt1979complexity}. However, precision locking avoids this problem by directly evaluating predicate against before- and after-images of posted writes~\cite{jordan1981precision}. The technique was not used in any real system for a long time, but was later adopted by Neumann et al. for implementing the MVOCC mechanism of Hyper~\cite{neumann2015fast}. 

  

\section{Experimental Evaluation}
In this section, we present a comprehensive experimental evaluation of TreeCat. We set up a cluster of 4 physical machines on CloudLab~\cite{cloudlab}, each equipped with a 10 core Intel Xeon Silver 4114 processor, 192GiB DDR4 memory, and a 64GB SATA SSD, interconnected by 10Gbps ethernet. We implemented a catalog API, including $getTable()$ and $listFiles()$ operations, in Apache Spark 3.4, allowing communication with a standalone instance of TreeCat. 

\subsection{Concurrency Control Mechanism}
To evaluate the concurrency control mechanism, we implemented two standard concurrency control protocols in TreeCat. The first is OCC with optimistic scan range locking (OSL), which also primarily utilizes prefix-based scan range locks. However, instead of precision locking, it uses more fine-grained range locks for bounded range scans. For example, if only table partitions 1 to 100 are scanned, only the range of object id, $[1,100]$, rather than the entire range of table partitions, is optimistically locked. The second scheme implements S2PL with multiple granularity locking (MGL). All three schemes efficiently handle read-only transactions with an identical MVCC protocol. Our new proposed concurrency control mechanism, i.e., TreeCat's default protocol, is referred to as OSPL (optimistic scan range and precision locking).

We used a microbenchmark largely derived from the TPC-DS benchmark. While TPC-DS proceeds as a single fixed sequence of phases, we design our
benchmark so that multiple client threads randomly select queries and perform the corresponding metadata operations (against TreeCat) concurrently.
Besides the original read queries and data maintenance operations, which involve data insertion and deletion on the fact tables, we introduce two new
operations: an \textbf{optimize} operation, which selects and merges small files, and \textbf{dimension table insertion}. These new operations allow
the simulation of real-world scenarios that require support for {\bf multi-table transactions}. The relative distribution of write operations is adjusted with realistic assumptions about standard data warehouse workloads. For example, we assume that fact table insertion occur 12 times more frequently than dimension table insertion (every 5 minutes vs. every hour) and 144 times more frequently than optimize operations (every 5 minutes vs. every 12 hours). We adopted the schema of the TPC-DS benchmark, but used a custom data generator to generate relatively large amount of metadata. The amount of metadata per table, the cardinality of surrogate keys and business ids are in line with the TPC-DS specification for a scale factor of 100 terabytes. In our experiment, we deployed TreeCat on a single machine and client threads on the other three machines.

We first conducted a scalability experiment to measure overall throughput, abort rate, and latency over a set time duration across different numbers of client threads and percentages of read-write transactions. For all three schemes, read-only transactions do not cause any conflict as they follow the MVCC protocol. Consequently, performance differences between the schemes are minimal for more read-intensive workloads, as shown in \autoref{fig:concurrency1}. However, for write-intensive workloads, throughput diverges, with OSPL significantly outperforming the other two as the number of client threads increases. The main source of contention is read-write conflicts on the customers table, which is one of the dimension tables to which data is inserted periodically. Inserting into a fact table requires joining the input data with dimension tables on the business ids (primary key in the source OLTP database) and should only conflict with dimension table insertion if the ranges of dimension table business IDs overlap. However, both OSL and MGL apply coarse-grained scan range lock on the dimension tables, leading to unnecessary conflicts. As a result, the abort rate of OSL increases almost linearly with the number of threads, reaching 31.3\% for the balanced workload and 63.7\% for the write-intensive workload at 30 client threads. Similarly, MGL maintains low throughput since transactions are blocked while waiting for lock acquisition. By contrast, OSPL avoids false conflicts through precision locking on the file statistics, achieving significantly higher throughput and a low abort rate. The latter two plots of \autoref{fig:concurrency1} show this trend more clearly as the percentage of read-write transactions increases for 30 client threads. A considerable proportion of read-only transactions scan entire fact tables and have longer latency than read-write transactions, which is why the throughput of OSPL increases with more write-intensive workloads.  

\begin{figure}[ht]
  \centering
  \includegraphics[width=\linewidth]{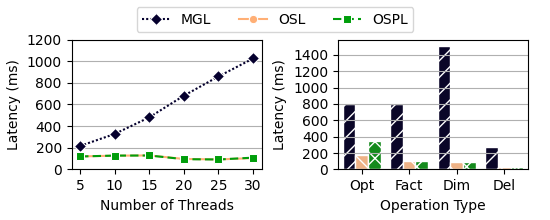}
  \caption{99th percentile operation latency of read-write transactions across different number of client threads and operation types for 30 threads, respectively. In the latter figure, we use abbreviations: Opt (optimize), Fact (fact table insertion), Dim (dimension table insertion), and Del (delete).}
  \label{fig:concurrency2}
\end{figure}

Next, we compare latency across the different schemes for the balanced workload. Because read-only operations have longer latencies than read-write transactions and follow the same MVCC protocol in all three schemes, we isolate and analyze the latencies of read-write transactions. \autoref{fig:concurrency2} shows the 99th percentile latency of read-write transactions across different number of threads and for each operation type. While the latency remains consistent for the two OCC schemes, it increases almost linearly with the number of client threads for MGL as transaction lock queues build up. Categorizing transactions by operation type reveals that the latency of dimension table insertion is almost as twice as long as that of fact table insertion, indicating that the conflicts on the updated dimension table is the main source of contention for MGL. {\em The latency of optimize operation is significantly longer for OSPL compared to OSL because higher number of small files accumulate with less aborts on insertion, resulting in larger optimize operation.}

\begin{figure}[ht]
  \centering
  \includegraphics[width=\linewidth]{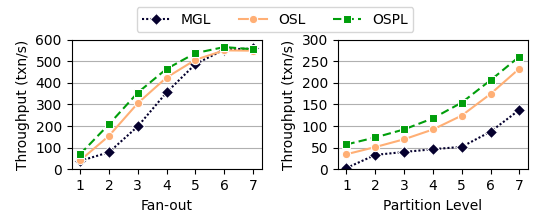}
  \caption{Throughput for different breadths (fan-out) and depths (partition level) of the metadata schema.}
  \label{fig:breadthdepth}
\end{figure}

Lastly, we evaluate how two important structural properties of the metadata schema, breadth and depth, affect performance. These experiments use a separate microbenchmark designed to be more generalizable. The workload consists of a mix of read and read-write operations (50-50) on a single data set, which is range-partitioned into 100000 data files by a set of independent clustering attributes. Both types of operation perform a predicate read operation with fixed selectivity, using range predicates on randomly selected clustering attributes. Read-write operation additionally inserts a new data file. A partition level is constructed by grouping data files by a clustering attribute.  Nested grouping by multiple attributes results in a multi-level hierarchy where the number of such attributes determines the overall partition level (depth) and the cardinality determines the fan-out (breadth). As shown in \autoref{fig:breadthdepth}, throughput is low across all schemes at low fan-out because of large scan ranges (e.g., all 100000 files have to be scanned at the partition level of 1). However, both OSL and MGL are further affected by contention from coarse-grained scan range locking. OSPL, on the other hand, avoids false conflicts as it uses a predicate-based method. Throughput converges with higher fan-out, as MGL and OSL benefit from more fine-grained lock ranges. A similar trend can be observed with different partition levels (depths) for the same reason. However, throughput is divergent even at the highest partition level because a subset of the clustering attributes do not form partitioning groups; a predicate read operation only on these attributes requires scanning all partition groups, resulting in higher contention for OSL and MGL. Although OSL outperforms MGL in throughput, it suffers from high abort rates.

\subsection{Comparative Evaluation}
Next, we compare TreeCat with HMS, Delta Lake, and Iceberg. An apples-to-apples comparison is difficult because TreeCat does not support a catalog API fully compatible with Spark and all the systems have different architectures and transactional guarantees. We nonetheless try our best to compare the relative costs of metadata operations and conduct an end-to-end experiment with an alternate database engine. We set up HDFS (Hadoop Distributed File System) and Spark on the 4 node cluster. We used Hive 2.3, Delta Lake 2.4, and Iceberg 1.5, all of which are compatible with Spark 3.4. We used a single table with a schema identical to that of the Store Sales table in the TPC-DS, partitioned by the sales date, from 1998-01-01 to 2003-12-31. To test scalability of read performance, we populated the table with varying numbers of files and measured the latency of retrieving files with a range filter on the date. While our primary implementation of the catalog API uses a single thread to retrieve the query result via a single stream, both HMS and Iceberg leverage multiple cores on the client machine to retrieve the result over multiple streams where HMS calls $listFile()$ and Iceberg retrieves manifest files in parallel. Client-side parallelism can be achieved with TreeCat quite easily by adding a partitioning predicate to the path query, so that multiple threads can concurrently submit queries and retrieve distinct result sets. Our implementation applies the $endwith()$ operation to the partition value due to the limitations of the current query language. We used this parallel implementation for the following experiments. 

\begin{figure}[h]
  \centering
  \includegraphics[width=\linewidth]{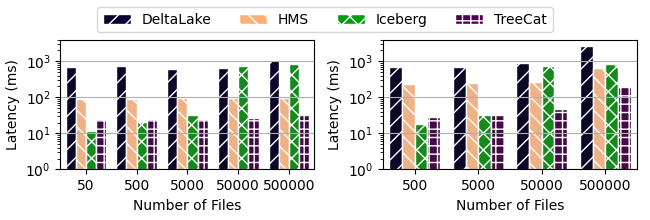}
  \caption{Median latency of file retrieval across different number of files for 1 day and 365 days predicate ranges, respectively.}
  \label{fig:scan1}
\end{figure}

As shown in \autoref{fig:scan1}, we first analyze the scalability of the file retrieval operation with high selectivity (date range of 1 day) and with low selectivity (date range of 365) across different numbers of files. For the 1-day predicate range, TreeCat consistently shows low latency as it efficiently traverses the hierarchical data on the server and only returns the file metadata in the given data partition. HMS also scales quite well as it calls $listPartitionsbyFilter()$ to only retrieve the partitions in the given range from the thrift server. Although Delta Lake benefits from Spark's distributed processing engine, the overhead of executing the job adds significant latency, resulting in a relatively high minimum latency of around 600 ms. This contrasts with previous experimental results~\cite{armbrust2020delta,jain2023analyzing} and we suspect that performance could be significantly improved with a better-configured, larger Spark cluster. But the client-side implementation of getting the end result appears to be a major bottleneck. Although Iceberg utilizes multiple cores for processing manifest files, the manifest file size can get large, resulting in the longer latency for the higher file counts. TreeCat outperforms the other systems and scales well also for the 365-days predicate range. HMS beats both Delta Lake and Iceberg when the file count reaches 50000, which we mainly attribute to the low latency of HDFS.

\begin{figure}[t]
  \centering
  \includegraphics[width=\linewidth]{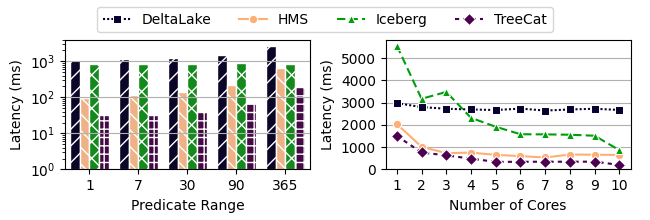}
  \caption{Median latency of file retrieval across different predicate ranges and numbers of physical cores on client machine, resp. File count is 500000 for both while predicate range is 365 for the latter.}
  \label{fig:scan2}
\end{figure}

Next, we analyze how the selectivity of the date range filter affects the latency for the file count of 500000. Both TreeCat and HMS effectively prune irrelevant date partitions, resulting in better performance for narrower predicate ranges. Iceberg, on the other hand, operates at the granularity of manifest files, so the latency remains relatively high. Delta Lake also does not scale too well as the number of retrieved files increases. We also evaluate how the number of physical cores on the client machine affects the overall performance, using 365-day date range and a file count of 500000. As expected, the performance of HMS, Iceberg, and TreeCat, which take advantage of client-side parallelism, degrades with limited number of cores while that of Delta Lake remains consistent.

\begin{figure}[t]
  \centering
  \includegraphics[width=0.99\linewidth]{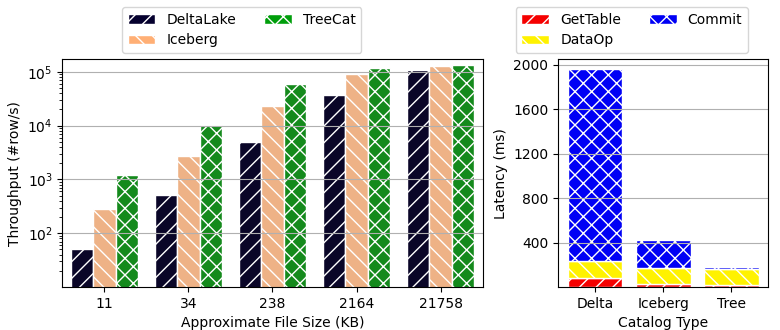}
  \caption{Throughput of insert operations (log scale) for different write sizes, and time breakdown for an insert operation of
      10000 row data file ($\approx$ 238 KB), respectively.}
  \label{fig:endtoend}
\end{figure}

We conclude with an end-to-end experiment, using DuckDB 1.2 as the execution engine. The workload involves repeatedly inserting data of constant size into an unpartitioned Store Sales table. A single operation proceeds in three steps: (1) Retrieve the table metadata, including the table directory and schema from the catalog; (2) Parse the schema, generate data, and write the output to a parquet file under the table directory, using DuckDB; (3) Invoke the commit protocol on the catalog with the metadata of the new data file. We measure both the total throughput over a set duration and the end-to-end latency of each invocation across different data file sizes, configured using the number of rows per file (powers of 10, starting from 100). The results are shown in \autoref{fig:endtoend}. We observe that the high latency of metadata operations can indeed cause a performance bottleneck, especially for smaller file sizes. As shown in the time breakdown of \autoref{fig:endtoend}, the main difference lies in the commit protocol: Delta Lake's commit protocol takes over 1.5 seconds, and Iceberg takes around 250 ms, while TreeCat only takes around 10 ms. The commit protocol for lakehouse storage formats involves retrieving the latest table snapshot (which may require reading and processing metadata files), writing one or more metadata files, and invoking a CAS (compare-and-swap) operation for ACID guarantees. The entire process involves one or more operations on the shared storage system, which have non-trivial overhead. Delta Lake incurs extra overhead from launching a Spark job for metadata processing. In contrast, TreeCat uses in-memory data structures for validation, applies writes to RocksDB, which is optimized for fast write, and is only delayed by {\em fsync()} operation on the WAL file, resulting in lower latency. The throughput eventually converges with larger file sizes as the cost of the data operation becomes more dominant. However, given applications such as streaming systems~\cite{carbone2015apache,armbrust2018structured} that generate small data files at high frequency, these results carry significant real-world implications.

\section{Conclusion and Discussion}
\label{sec:discussion}

In this paper, we identified the primary use cases of a standalone operational catalog and justified the key design decisions for our catalog engine, TreeCat. Our experimental evaluation not only exposes the limitations of existing solutions but also validates the advantages of our design choices.


Looking ahead, we see a number of critical challenges in this space that need further research.
Some of these challenges lie outside the scope of the catalog itself. For instance, {\bf consistency between the data and the
metadata} (e.g., ensuring that a pointer to a physical object is valid), is a major emerging challenge. Yet it is not clear where that responsibility
lies. 
Similarly, {\bf access control} is crucial but needs to be enforced by the layer that manages access to the data itself. {\bf Garbage collection} of data objects and their metadata that are no longer visible may become necessary over time. A TreeCat instance can be 
periodically scanned to find and return leaf objects that are no longer reachable by any path from the root. These objects can be quarantined in
temporary storage until the data layer invokes a final commit to remove them. Finally, {\bf scalability and availability} are important concerns in a distributed setting. Given the relatively small volume of metadata and large numbers of
read queries, we propose the adoption of a single-writer, multiple-reader architecture where all read-write transactions are handled by a single server, and logs are shipped to and replayed by read replicas ~\cite{verbitski2017amazon}. This scheme would require minimum changes to the concurrency control mechanism because a single server will process all write operations. If the primary writer server fails, a new writer server can be elected via a quorum-based mechanism, ensuring
 availability. 

\begin{acks}
 This research was supported in part by Dolby Laboratories.
\end{acks}


\balance
\bibliographystyle{ACM-Reference-Format}
\bibliography{ref}

\end{document}